\documentclass[aps,twocolumn,amsmath,amssymb,prl,superscriptaddress,preprintnumbers,nofootinbib,10pt]{revtex4-1}
\usepackage{graphicx,subfigure}
\usepackage{pictexwd,dcpic}
\usepackage{color}

\begin{document}
\title{Classical de Sitter Solutions of Ten-Dimensional Supergravity}
\author{Clay C\'{o}rdova}
\affiliation{School of Natural Sciences, Institute for Advanced Study, Princeton, NJ 08540, USA}
\author{G.~Bruno De Luca}
\author{Alessandro Tomasiello}
\affiliation{Dipartimento di Fisica, Universit\`a di Milano--Bicocca, Piazza della Scienza 3, I-20126 Milano, Italy and \\ INFN, sezione di Milano--Bicocca, I-20126 Milano, Italy}

\begin{abstract}
\noindent We find four-dimensional de Sitter compactifications of type IIA supergravity by directly solving the ten-dimensional equations of motion. In the simplest examples, the internal space has the topology of a circle times an Einstein manifold of negative curvature. An orientifold acts on the circle with two fixed loci, at which an O8$_-$ and an O8$_+$ plane sit. These orientifold planes are fully backreacted and localized.  While the solutions are numerical, the charge and tension of the orientifold planes can be verified analytically. Our solutions have moduli at tree level and can be made parametrically weakly-coupled and weakly-curved.  Their fate in string theory depends on quantum corrections. 
 
\end{abstract}
\maketitle

Modeling dark energy is a basic challenge for fundamental physics. For theories of gravity based on extra dimensions it is notoriously difficult to obtain positive cosmological constant (or ``de Sitter'') compactifications  \cite{gibbons-nogo}.  In particular,  in the supergravity theories that describe string theory at low energies, de Sitter solutions are impossible if one only makes use of the two-derivative classical action and ingredients obeying standard energy conditions \cite{dewit-smit-haridass,maldacena-nunez}. 

To evade these constraints it is typical to invoke either quantum corrections to the effective action, or to use classical objects, notably orientifold-planes (O-planes), that have negative tension. Over the years, several classes of de Sitter models have been proposed using these ideas (see for example \cite{kklt,balasubramanian-berglund-conlon-quevedo, silverstein-simple, caviezel-koerber-kors-lust-wrase-zagermann}).

Most of the de Sitter solutions to date make use of a four-dimensional effective theory to describe the physics of the vacuum. When used consistently, such effective theories can be a helpful tool.  However, the non-linear nature of gravity makes it difficult to rigorously justify these four-dimensional approximations.  Moreover, typical constructions involve many ingredients, and inevitably some doubts have lingered even over seemingly robust constructions \cite{bena-grana-halmagyi, banks-10500, Sethi:2017phn}, with some voices even raising skepticism on the very existence of de Sitter vacua in string theory \cite{danielsson-vanriet,obied-ooguri-spodyneiko-vafa,Garg:2018reu, ooguri-palti-shiu-vafa}.

The most conservative and robust approach to address these concerns is to look for simple classical de Sitter solutions of the full ten-dimensional supergravity equations.   This classical regime is justified provided that we can find solutions that exist at small string coupling and curvature so that all quantum corrections can be ignored.  Localized negative energy is then provided by orientifold planes.\footnote{In string theory orientifold planes come in two main varieties O$_{+}$ and O$_{-}$.  They have respectively positive and negative tension, and opposite Ramond--Ramond charge.  The worldvolume theory of D-branes coincident with an O$_{-}$ has an $\mathrm{SO}$ gauge group, while D-branes coincident with O$_{+}$ give $\mathrm{Sp}$ gauge groups.  Our solutions below will involve both types of O8-planes. Another configuration involving both types of O8-planes is the one considered in \cite{dabholkar-park,witten-without,aharony-komargodski-patir}, dual to M-theory on a Klein bottle.}

The supergravity backreaction one should expect for O-planes is known from flat-space solutions, but until recently few nontrivial compactifications including them were known. As a fallback strategy, some progress was obtained \cite{danielsson-haque-shiu-vanriet,danielsson-koerber-vanriet} by pretending that the O-planes could be ``smeared'', namely distributed smoothly over the internal manifold. The solutions obtained this way are encouraging, but are not physical, since O-planes have to be located at fixed points of involutions and are non-dynamical.\footnote{Recent attempts with more exotic strategies include \cite{decarlos-guarino-moreno,soueres-tsimpis,harribey-tsimpis, Banerjee:2018qey}. } 

More recently, however, compactifications with internal localized O-planes have started to appear; for example \cite{brandhuber-oz,afrt,bah-passias-t}. Their backreaction gives rise to strong curvature and string coupling near the O-plane, and hence to a localized breakdown of supergravity. At first sight this may look discouraging.  However, the orientifold is an exact (supersymmetric) solution of string theory whose local behavior is correctly reproduced in these constructions.  Moreover, since these examples have negative cosmological constant, they can be tested using holography \cite{cremonesi-t,bah-passias-t}, and this gives further confidence that the O-plane backreaction is in fact under control.

Encouraged by these results, we now present a new class of de Sitter compactifications to four dimensions from massive type IIA supergravity, using O8-planes (O-planes with eight space dimensions) which are backreacted and correctly localized.\footnote{A similar construction in an effective $5d$ theory appeared in \cite{EvaStrings}.  See also \cite{Andriot:2016xvq} for some constraints on de Sitter solutions with O8s.}  

Our simplest class of solutions are warped products dS$_4\times M_6$, with metric\footnote{Throughout we work in string units.}
\begin{equation}\label{eq:metriceasy}
	ds^2_{10}= e^{2W} ds^2_{\mathrm{dS}_4}+  e^{-2W}\left(dz^2+e^{2\lambda} ds^2_{M_5}\right)\,.
\end{equation}
In the above,  $ds^2_{\mathrm{dS}_4}$ is a metric on de Sitter space with cosmological constant $\Lambda$.  Meanwhile, the geometry of the internal space is such that $z$ is periodically identified and $M_{5}$ is an Einstein manifold with constant Ricci scalar $5\kappa$.  The overall scale of $\kappa$ is unphysical; it can be absorbed by shifting $\lambda$.  However,  we will presently see that $\kappa<0$ follows from the equations of motion. We take the functions $W,$ $\lambda$, as well as the dilaton $\phi$ to be functions of the coordinate $z$ alone.  Finally, the only flux present is the Romans mass $F_{0}$.

Our analysis parallels \cite{cordova-deluca-t-ads8}, where a similar ansatz was used to find AdS$_8$ solutions. After some manipulation, the equations of motion away from sources read:
\begin{subequations}\label{eq:easy}
\begin{align}
  0&=\frac{25 \kappa  e^{-4 \phi /5}}{\alpha ^{2/5}}+\frac{4 \alpha' \phi'}{\alpha }-\frac{4
   \left(\alpha'\right)^2}{\alpha ^2}-\frac{5}{2} F_0^2 e^{2 \phi -2 W}  \label{eq:easy1} \\&-20 W' \phi '+40
   \left(W'\right)^2+20 \Lambda  e^{-4 W}+4 \left(\phi '\right)^2~, \nonumber \\
 0&=F_0^2 e^{2 (W+\phi )}+4 \Lambda -4 e^{4 W} W''-\frac{4 e^{4 W} \alpha' W'}{\alpha } ~,\label{eq:easy2}\\
 0&=\alpha  \Lambda +\alpha ^{3/5} \kappa  e^{4 W-\frac{4 \phi }{5}}-\frac{1}{5} e^{4 W} \alpha''~,
 \label{eq:easy3}
\end{align}
\end{subequations}
where primes indicate derivatives with respect to $z$ and we have introduced the function
\begin{equation}
\alpha \equiv e^{5\lambda - 2\phi}\,.
\end{equation}

As mentioned above, the sources in our solutions are O8-planes.  (In footnote \ref{foot} below we also comment on the possibility of including D8-branes).  We take them to be extended along all directions except $z$. At the locus $z=z_0$ of one of these sources the functions describing the ansatz are continuous, but have discontinuities in their first derivatives that are fixed by the equations of motion:
\begin{subequations}\label{eq:dweasy}
\begin{align}
	&e^{W-\phi}\Delta W' = \frac15 e^{W-\phi}\Delta \phi' = -\frac{1}{4} \Delta F_0 \, , \ \ \  \Delta \alpha' = 0~,\\
	&	F_0|_{z\to (z_0)_-} =-e^{W-\phi} \left(\phi'- W' + \frac{\alpha'}\alpha\right)\;. \label{eq:stop}
\end{align}	
\end{subequations}
We also have that $-2\pi\Delta F_0$ is the charge of the source: this equals $1$ for a D8. For the two types of O8-planes, called O8$_\pm$, the charge is $\pm 8$.

\begin{figure}[ht!]
	\centering
		\includegraphics[width=6cm]{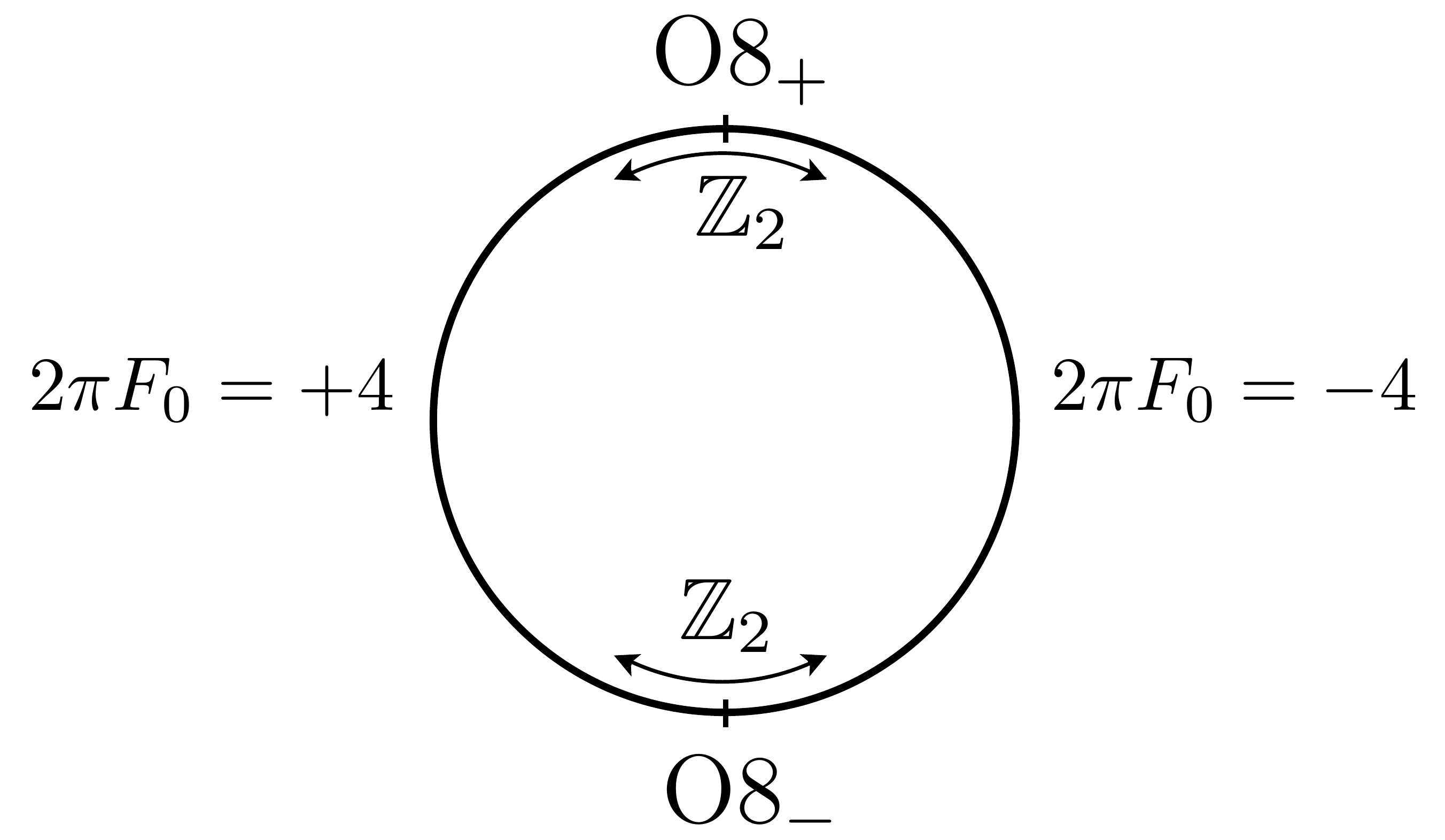}
	\caption{A pictorial representation of the coordinate $z$ together with the $\mathbb{Z}_2$ orientifold action resulting in two O8-planes. }\label{fig:O8Z2}
\end{figure}

In our de Sitter solutions, the coordinate $z$ is periodically identified according to $z\sim z+ 2 z_0$, and there is an orientifold action $z \to -z$.  At the fixed locus $z=0$ there is an O8$_+$, while at $z=z_{0}$ there is an O8$_{-}$ (see Figure \ref{fig:O8Z2}).  Correspondingly we have $F_0=-\frac 4{2\pi}$ for $z\in (0,z_0)$, while in the mirror region, $z\in(-z_0,0)$, $F_0= \frac 4{2\pi}$. All the metric coefficients are even under $z\to -z$.

We can see a constraint on the parameters of our ansatz by evaluating \eqref{eq:dweasy} for small positive $z$ tending towards zero. Approaching from this direction we find  $W'(0)=\frac{1}{5} \phi'(0)=\frac1{2\pi} e^{\phi-W}(0)$, $\alpha'(0)=0$. Imposing this on the first order equation \eqref{eq:easy1} we then find that 
\begin{equation}\label{lambdasolve}
\Lambda = -\frac{5}{4}\kappa e^{4W-2\lambda}~,
\end{equation} where the functions are evaluated on the O8$_+$. (A similar formula appeared in \cite{gautason-junghans-zagermann}.) As promised, for $\Lambda>0$ we see that the scalar curvature $\kappa$ must be negative.  In the equations below we set $\kappa=-\frac{4}{5} \Lambda$ for convenience.

In the vicinity of the O8$_{+}$ (or other finite coupling sources) one can readily solve the equations of motion \eqref{eq:easy} perturbatively.  At the first few orders we obtain: 
\begin{subequations}\label{eq:localD8easy}
\begin{align}
e^{-4W} &=c_1+\frac{F_0 }{\sqrt{c_2}}z-2 c_1^2 \Lambda z^2+\mathcal{O}\left(z^3\right)~,  \\
e^{-\frac{4}{5} \phi} &=c_1 c_2^{2/5}+\frac{F_0 }{\sqrt[10]{c_2}}z-\frac{c_1 F_0 \Lambda }{6
   \sqrt[10]{c_2}}z^{3}+\mathcal{O}\left(z^4\right)~,\\
\alpha &= c_2+\frac{c_1 c_2 \Lambda}{2}   z^2+\frac{\sqrt{c_2} F_0 \Lambda}{6}   z^3+\mathcal{O}\left(z^4\right)~. 
\end{align}
\end{subequations}
Here, $c_1$, $c_2$ are constants parametrizing the solution, and $F_{0}$ is as specified in Figure \ref{fig:O8Z2}.

As a consistency check, one can compare the above to an orientifold in flat space.  This takes the familiar form
\begin{equation}\label{eq:O8easy}
	ds^2_{10} \sim H^{-1/2}(-dx_0^2 + dx_1^2 +\ldots dx_8^2)+ H^{1/2} dz^2\,,
\end{equation} 
and $e^\phi \sim g_s H^{-5/4}$, where $H$ is a harmonic (i.e.~linear) function of $z$.  Using \eqref{lambdasolve} we identify $H\sim e^{-4W}$ and for $\Lambda\rightarrow 0$ in \eqref{eq:localD8easy} the perturbative expansion truncates leading to the expected behavior. 

The local solution (\ref{eq:localD8easy}) is valid for $z>0$, and the functions for $z<0$ are defined to be even under $z\to -z$. The first derivatives then are discontinuous as required in (\ref{eq:dweasy}). Thus the solution has a singularity at $z=0$, as expected from the presence of an O8$_+$. This singularity, however, is milder than the ones displayed by orientifolds of lower dimensionality, where fields usually diverge.

\begin{figure}[ht!]
	\centering
		\includegraphics[width=6.5cm ]{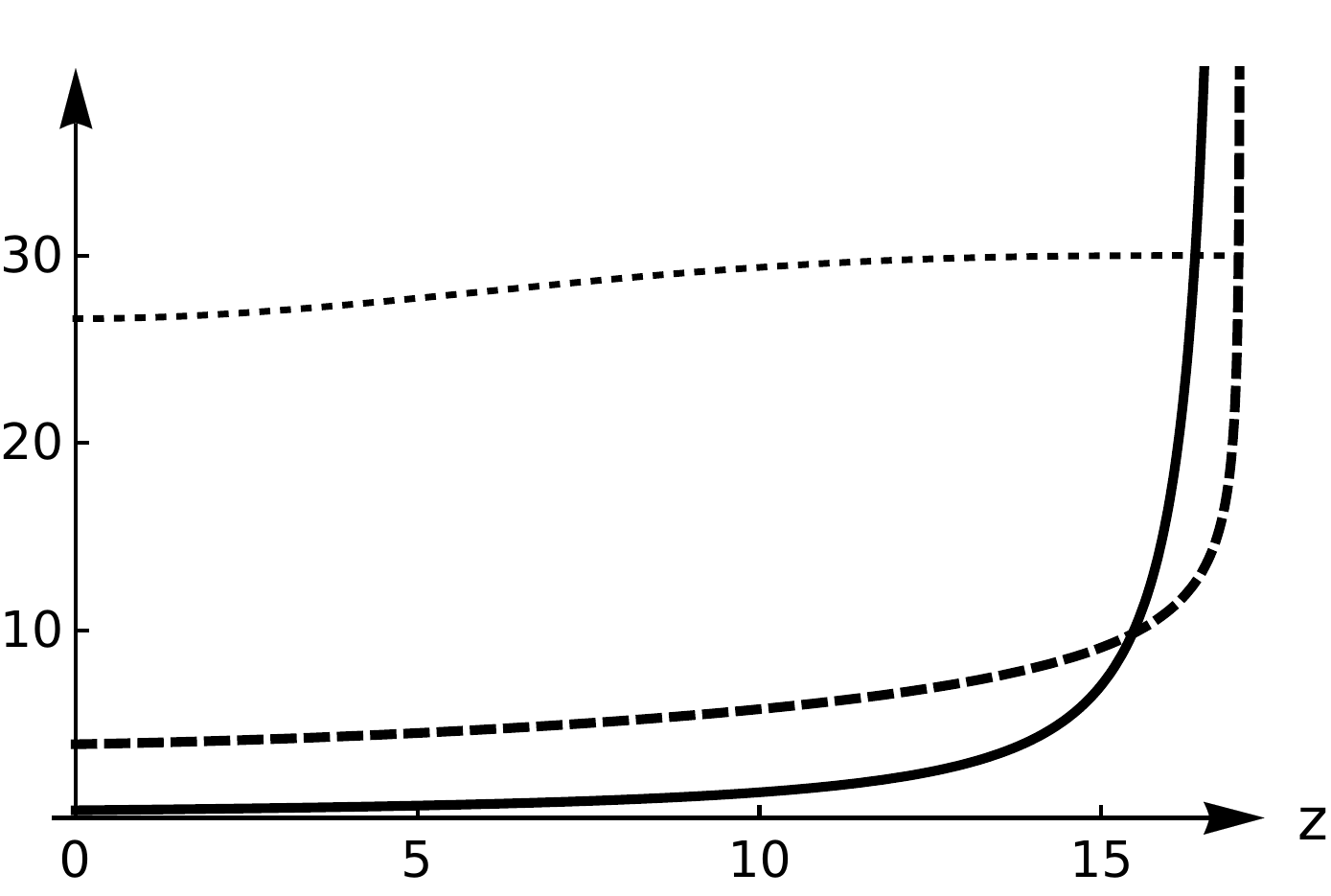}
	\caption{A numerical solution with $\Lambda = 1$.
The functions are $e^\phi$ (solid), $e^{W}$ (dashed), $\alpha$ (dotted, rescaled). At the right endpoint, it behaves as an O8$_{-}$ with diverging dilaton.}\label{fig:ds4D8O8}
\end{figure}

One can now evaluate the solution \eqref{eq:localD8easy} at a small $z$ and start a numerical evolution. For a certain open set in the space of the constants $c_i$, the solution ends with a singularity which we recognize numerically to be locally identical to the O8$_-$ with diverging dilaton.  This again takes the form \eqref{eq:O8easy}, but now $H= \frac{g_s}{4\pi}|z|$. Moreover, one can check that on a numerical solution with this behavior, the equations (\ref{eq:dweasy}) are automatically satisfied, except for $\alpha'(z_0)= 0$, which can be arranged with a one-parameter fine-tuning of the initial conditions. Figure \ref{fig:ds4D8O8} illustrates a typical numerical solution. 

Near the O8$_-$ one can also perform a perturbative analysis like that resulting in \eqref{eq:localD8easy}. This is done by imposing that the leading power behavior for the dilaton and metric coefficients is the one inferred from \eqref{eq:O8easy}. On the resulting local solution (which matches with our numerical one near the O8$_-$) all the conditions in \eqref{eq:dweasy} are automatically satisfied, with the correct tension.

An important caveat to our solutions (and other supergravity backgrounds with orientifolds) is that at the orientifolds the curvature becomes large and the supergravity approximation is locally not valid.  This is especially so in our examples where the string coupling is also large near the O8$_-$.  This makes the derivation of the boundary conditions (\ref{eq:dweasy}) sensitive to string corrections, and even formally their extrapolation to infinite coupling at the O8$_-$ is ambiguous due to this divergence (see \cite{cordova-deluca-t-ads8} for a similar discussion).  Nevertheless, the fact that the system is attracted to solutions with this behavior, which is the same for an O8$_-$ in flat space, makes us think that (\ref{eq:dweasy}) are still physically relevant, and that they are a good way to ensure the correct tension for the orientifolds.

The simple class of de Sitter solutions presented above can be enlarged by including additional fluxes.  For instance, we can generalize our metric ansatz to 
\begin{equation}\label{eq:metrichard}
	ds^2_{10}= e^{2W} ds^2_{\mathrm{dS}_4}+  e^{-2W}\left(dz^2+e^{2\lambda_2} ds^2_{M_2}+e^{2\lambda_3} ds^2_{M_3}\right)\,.
\end{equation}
Compared to our previous example, we have split the five-manifold $M_{5}$ into $M_2$ and $M_3$ which are two Einstein spaces with Ricci scalars $2 \kappa_2$ and $3 \kappa_3$; below we see that at least one of them must be negative.  Again we take $W,$ $\lambda_{i}$, $\phi$ to only depend on $z$, and in addition to $F_0$ we allow
\begin{equation}\label{eq: defF4}
F_4 = f_4 e^{-6W+3\lambda_3-2\lambda_2}dz\wedge \text{vol}_{M_3}~.
\end{equation}
Here $f_4$ is a constant, and the $z$ dependence has been chosen such that the equation of motion $d\star F_4  = 0$ is automatically satisfied.

In this more general setup, the equations of motion away from sources read:
\begin{subequations}\label{eq:eqD1}
\begin{align}
0&=-F_0^2e^{2 \phi -2 W}+6 \kappa _3 e^{-2 \lambda _3}-12 \lambda _3' \phi '+15
   \left(\lambda _3'\right)^2\label{eq:eqD1-1}\\
 +&2\frac{\alpha'}{\alpha}\left(-3  \lambda _3'+2  \phi '\right)+\frac{4 \kappa _2 e^{3 \lambda _3-2 \phi }}{\alpha
   }-\left(\frac{\alpha '}{\alpha}\right)^2+4 \left(\phi
   '\right)^2\nonumber\\
&-8 W' \phi'+16 \left(W'\right)^2+8 \Lambda  e^{-4 W}+f_4^2 \frac{ e^{6 \lambda _3-6 W-2 \phi }}{\alpha
   ^2}~, \nonumber
\end{align}
\begin{align}
0&=\Lambda-e^{4 W} \left(W''+\frac{\alpha' W'}{\alpha }\right)\label{eq:eqD1-2} \\
&+\frac{f_4^2 e^{6 \lambda _3-2 (W+\phi )}}{ 4\alpha ^2}+\frac{F_0^2}4 e^{2 (W+\phi )} ~, \nonumber
\end{align}
\begin{align}
 0&=-\frac{\alpha''}{\alpha}+2 \kappa _2\frac{ e^{3 \lambda _3-2 \phi }}{\alpha }+5 \Lambda  e^{-4 W}\label{eq:eqD1-3}\\
&+\frac{f_4^2 e^{6 \lambda _3-6   W-2 \phi }}{2 \alpha ^2}+3 \kappa _3 e^{-2 \lambda _3}~,\nonumber
\end{align}
\begin{align}
 0&=-\frac{\alpha''}\alpha +2 \kappa_2 \frac{e^{3 \lambda_3-2 \phi}}{\alpha}-2 \Lambda  e^{-4 W}\label{eq:eqD1-4}\\
&+ \frac{\alpha'}{\alpha} \left(3\lambda _3+4  W-2 \phi  \right)'+\left(3 \lambda _3+4 W-2 \phi \right)''~,\nonumber
\end{align}
\end{subequations}
where now $\alpha \equiv e^{2\lambda_2+3\lambda_3 - 2\phi},$ and at the sources we must also provide the discontinuity equation $\frac12 \Delta \lambda_3'=\Delta W'$.

Solutions can now be constructed as above.  We begin with a finite coupling O8$_{+}$ at $z=0$ and evolve to an infinite coupling O8$_{-}$ at $z=z_{0}$. Note that $F_{4}$ in \eqref{eq: defF4} is odd under the orientifold as expected.  In this case, the analog of the constraint \eqref{lambdasolve} is
\begin{equation}
\Lambda = -\frac{1}{2} \kappa _2 e^{-2 \lambda _2+4W}-\frac{3}{4} \kappa _3 e^{-2 \lambda _3+4W}-\frac{f_4^2}{8}e^{-4 \lambda _2-2 W+2 \phi }~,
\end{equation}
where the right-hand-side is evaluated at the O8$_{+}$.  From this we see in particular that at least one of the $\kappa_{i}$ must be negative.  The resulting solutions depend on two continuous parameters, which can be thought of as the remaining initial conditions of the solution near the O8$_{+}$ after tuning to hit the O8$_{-}$.\footnote{As another consistency check of these equations, we can see that they admit a solution of the form AdS$_4\times \mathbb{H}_2 \times S^4$, which is a simple variation on the AdS$_6\times S^4$ of \cite{brandhuber-oz}. This is most easily seen by going to a gauge where $dz^2$ in \eqref{eq:metrichard} is replaced by $e^{2Q} dz^2$. Then the solution is obtained by setting  $Q = 2 W$, $\lambda_2 = 2 W$, $\lambda_3 = 2 W + \log(\sin z)$, $\phi = 5W $+const.,  $W = -\frac16 \log(F_0 \cos z)$+const., $\kappa_2 = \Lambda $,  and $\kappa_3 = 2$. This results in a negative cosmological constant.}

Let us now comment on the properties of these de Sitter solutions. The first significant feature is that all our examples have classical moduli; i.e.\  the solutions come in continuous families.  The number of moduli apparent from our construction is easily seen by parameter counting.  The local solutions \eqref{eq:easy} depend on two continuous parameters and require a one-parameter tuning to reach a physical O8$_{-}$, resulting in one modulus.  The more general solutions \eqref{eq:metrichard} have two moduli.

One way to understand some of these moduli is that the equations of motion are invariant under the rescaling 
\begin{equation}
g_{MN}\to e^{2c} g_{MN}~,~~~ \phi\to \phi-c~, ~~~F_{4}\to e^{4c}F_{4}~.
\end{equation}
This rescaling can be used to make the coupling and curvature as small as one wants, and in particular to parametrically reduce  the region around the O8$_{-}$ where supergravity breaks down. 

In the simplest solutions of type \eqref{eq:metriceasy}, the four-form flux vanishes and the single modulus is the parameter $c$ above.  In the more general solutions \eqref{eq:metrichard} with non-zero $F_{4},$ flux quantization implies that the rescaling parameter $c$ is discretized.  The two continuous moduli of these solutions do not admit such a simple presentation.

In the full string theory, one expects that quantum corrections will generate a potential on these moduli.  In the controllable regime of small string coupling these corrections are small, and might in principle be determined.  In the proximity of the O8$_-$, there is a localized breakdown of supergravity and all string corrections become relevant.  In general however, one no longer expects  \cite{dine-seiberg} the surviving vacua, if any, to be under parametrically good control.  For example, in the simplest class of solutions where $c$ appears to be a modulus in supergravity, we expect the potential to be such that $c$ is fixed in any true solution of string theory.

Within the classical supergravity approximation we can also try to analyze the stability of our solutions.  One obvious source of instabilities are tachyons.\footnote{\label{foot}Such Perturbative instabilities are present in related models we have studied involving D8 branes (either on top of the orientifolds or away from them).  In this case a probe computation as in \cite{cordova-deluca-t-ads8} shows that the position of the D8's along $z$ is tachyonic. } These can be assessed in our examples by a Kaluza--Klein (KK) reduction similar to that discussed in \cite{cordova-deluca-t-kk}.  Non-perturbative instabilities, such as brane bubble nucleation, should also be considered.

Let us also remark on the physical scales of these solutions.  By unwarping the metric, one can see that both the cosmological constant $\Lambda$ and the KK scale do not change as the modulus $c$ varies.   Thus there is no parametric separation of scales, although  it might be achieved by a suitable choice of the internal manifold.  By contrast, the effective 4d Planck constant scales as $e^{10c}$ and hence is large in the weak coupling regime.

In summary, we have found a new class of de Sitter solutions in supergravity. They are obtained directly from the classical equations of motion in ten dimensions, with fully localized orientifold-plane sources. Irrespective of the fate of these particular solutions, we expect our approach to be useful more broadly, for example for O$p$-planes for $p<8$, more general metric ans\"atze, and de Sitter solutions in other dimensions.\footnote{For instance, an identical analysis to the above leads to dS$_{d}$ solutions with only $F_{0}$ for any $2\leq d \leq 7$. }  Our strategy is conceptually very simple; it took some time to identify the right combination of ingredients and to build enough trust in the structure of O-plane singularities. Now that a first set of examples has been obtained, we expect more to follow.

\medskip\noindent {\bf Acknowledgements}
We thank O.~Bergman, D.~Junghans, J.~Maldacena, A.~Sagnotti, and T.~Wrase for discussions. CC is supported by DOE grant de-sc0009988. GBDL and AT are supported in part by INFN. 

\bibliography{at}

\end{document}